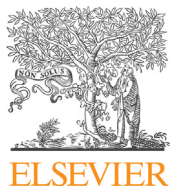
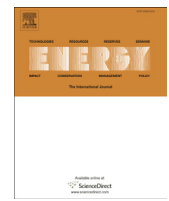

# Modeling all alternative solutions for highly renewable energy systems

Tim T. Pedersen [a, *], Marta Victoria [a], Morten G. Rasmussen [b], Gorm B. Andresen [a]

[a] *Aarhus University, Department of Mechanical and Production Engineering, Inge Lehmanns Gade 10, Aarhus C, 8000, Denmark*
[b] *Aalborg University, Department of Mathematics, Fredrik Bajers Vej 7K, Aalborg, 9220, Denmark*

A B S T R A C T

As the world is transitioning towards highly renewable energy systems, advanced tools are needed to analyze the complex energy networks. Energy system design is, however, challenged by real-world objective functions consisting of a blurry mix of technical and socioeconomic agendas, with limitations that cannot always be clearly stated. As a result, economically suboptimal solutions will likely be preferable. Here, a method capable of determining the continuum containing all economically near-optimal solutions is presented, moving the field of energy system modeling from discrete solutions to a new era where continuous solution ranges are available. The proposed method is applied to study a range of technical and socioeconomic metrics on a model of the European electricity system. The near-optimal region is found to be relatively flat allowing for solutions that are slightly more expensive than the optimum but better in terms of equality, land use, and implementation time.

© 2021 The Author(s). Published by Elsevier Ltd. This is an open access article under the CC BY license (http://creativecommons.org/licenses/by/4.0/).

## 1. Introduction

In the endeavor toward completing the ambitious climate goals of the Paris agreement, the energy supply of the world is under major revision to reduce carbon emissions [1]. Decarbonization of the global energy sector is a problem subject to large uncertainty [2], requiring extensive analysis when insights and guidance are to be provided for decision-makers [3]. An important tool used to study the possible futures for national and international scale energy supply systems is energy system optimization models (ESOMs). Based on historic or forecasted data on renewable resource availability, energy demands, network topology, and other technical constraints, the models optimize one or more objectives. Typically the primary objective is to reduce total system cost. A schematic of such a model is shown in Fig. 1. The modeled energy sectors often include electricity, and some models further include heating, transportation, gas, and biomass. Popular models of this type includes the technology rich MARKAL/TIMES model [4], the PRIMES model used by the European Commission to generate reference scenarios [5], and PyPSA [6] which is a open source model. A comprehensive review of popular ESOMs is available in Ref. [7].

Most of the ESOMs mentioned are linear optimization models with global objectives to minimize cost [8]. This formulation assumes that the cheapest solution is the best solution, however it could well be that society prefers a slightly more expensive solution that provides additional advantages such as decreased national/regional political disagreement [9], national and regional energy self-sufficiency [10], transition speed [11], public opinion [12] and technology unit size [13]. Many social and political incentives are often complex to represent mathematically or have blurry specifications and are therefore not included in models.

The limited ability for a model to represent all aspects of the real world is typically referred to as structural uncertainty as opposed to parametric uncertainty, which applies only to the data (parameters) used in the models [14]. As made clear by Pfenninger et al. [15], the quantification of uncertainties has become a key challenge for twenty-first-century energy system modelers. Parametric uncertainty can be addressed with a multitude of methods identifying the effect of changing input parameters on the model output. A thorough presentation of such methods is available in Ref. [16], with a classic example application on ESOM's seen in Ref. [17], where uncertainty related to temporal and cost data in a ESOM is investigated. Other applications of uncertainty analysis in ESOM's include analysis of uncertainties related to sector coupling between energy sectors [18], or exploring the effect of inter annual variability in weather data [19].

* Corresponding author.
  *E-mail address:* ttp@mpe.au.dk (T.T. Pedersen).

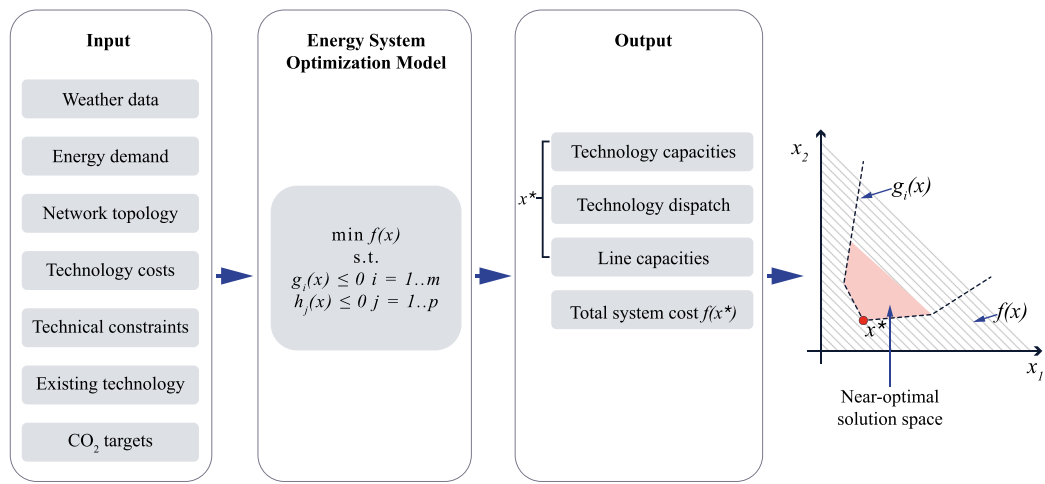

**Fig. 1. Energy system optimization model schematic** - The figure shows a schematic of an energy system optimization model. Typical input parameters used in ESOMs are shown on the left, combined with output variables on the right. The model is displayed in the most general mathematical form, with objective function $f(x)$, and constraints $g(x)$ and $h(x)$. The optimal solutions is given as $x*$. A two dimensional solution space is displayed on the right with the optimal solution marked as a red dot and the near-optimal solution space with marked with orange. (For interpretation of the references to colour in this figure legend, the reader is referred to the Web version of this article.)

The issue of structural uncertainty in energy system optimization models has, in recent years, gained strong attendance in the research community as researchers are starting to investigate solutions other than the optimum [20]. The exploration of other solutions than the economically optimal one is done using either multi-objective optimization [21], or methods such as the Modeling to Generate Alternatives (MGA) method [22]. Methods such as MGA can explore near-optimal solutions satisfying secondary objectives that cannot be clearly formulated in a linear mathematical expression, whereas multi-objective optimization requires a clear formulation of the secondary objectives. Furthermore, multi-objective optimization techniques are often limited to a handful of objectives, whereas MGA methods can explore problems of several hundred dimensions.

Most works addressing this issue share a common hypothesis arising from a paper by E. D. Brill from 1982 [23], where it is made clear that the real-world optimum cannot be determine due to inevitable uncertainties. Therefore, it is naive to study a single techno-economic optimal solution when it is most likely not the real-world optimum. Instead, the entire near-optimal solution space should be studied, as this set of solutions all represent possible real-world alternatives.

Several approaches have been proposed to study near-optimal solutions. Approaches range from the more indirect, where sensitivity analysis is conducted on constraints representing social acceptance issues [24]. Alternative approaches directly target the identification of near-optimal solutions. These approaches do so through iterative modification of the model objective function. The latter category of methods is known as Modeling to Generate Alternatives (MGA). The first MGA method was introduced in 1982 by Brill et al. [23] and then applied to energy system optimization models in 2011 by DeCarolis [25] is the so-called hop skip jump (HSJ) MGA method. It consists of three steps; (1) an optimal solution is found using the original model formulations. (2) a constraint is introduced, requiring that the value of the original objective function doesn't exceed that of the optimal solution plus a specified slack. (3) the objective function is altered to minimize the use of previously utilized technologies. Step 3 is then repeated until a desired number of alternatives are identified. The work by DeCarolis [25,26] has led to a range of papers using variations of the HSJ MGA method within the field of energy system optimization such as [27]: where the EXPANSE method is presented, and [28] presenting a method maximizing the difference between the found alternative solutions. Nacken et al. [29] combine the use of classic scenario based optimization with MGA for a model of Germany. Other variations of the MGA method, such as the one presented in Ref. [30], seek to find the boundary of the near-optimal solution space, thereby providing information about the minimum/maximum needed capacities of several technologies. Alternative approaches for generating near-optimal solutions use genetic algorithms such as the firefly algorithm [31], towards identifying alternative solutions within a given slack on total system cost [32].

Current state-of-the-art MGA algorithms, are based on the HSJ MGA algorithm presented in Ref. [23], but use more rigorous routines to explore near-optimal solutions. Sasse and Trutnevyte [9] find 100 random near-optimal solutions to decarbonize the power system for countries in central Europe divided into 650 regions. This enables them to quantify the social and environmental impacts of the near-optimal solutions at a regional scale. Neumann and Brown [30] implemented a One At the Time (OAT) approach, where the MGA objective function is chosen to iteratively maximize and minimize the capacity of every technology included in the model, one at the time. This ensures that the most extreme points of the near-optimal feasible space are identified. However, using the OAT approach in high dimensional problems may leave the majority of the uncertainty unexplored [33]. Lombardi et al. [34] followed a different approach in which they find 500 random near-optimal solutions, after which a small number of alternatives with the highest Euclidean spacing are chosen for further analysis. The common identifier for all available methods of addressing structural uncertainty in energy system models is that the result is a small finite set of alternative solutions. Having identified only a small number of alternative solutions, these methods fail in providing a robust mapping of the near-optimal feasible solution space.

Specifically, the problem with the use of the current state-of-the-art MGA algorithms can be summarized as follows: a) the dimensionality of the near-optimal solution space in energy system models is high. In state-of-the-art models, the dimension of the solution space is around the order $d \approx 10^2$ considering only investment variables and $d \approx 10^6$ considering investment and dispatch variables. To ensure sufficient coverage when sampling such complex spaces, very large sample sizes are required. Current MGA methods rely on very small sample sizes and provide no measures of convergence. Therefore, these methods must be improved to gain sufficient knowledge about the near-optimal solution space. b) Current methods provide no grantee for uniform coverage of the near-optimal feasible space. This introduces the possibility of bias towards certain solutions. c) Having identified only a small number of alternative solutions, it is not possible to extract information about variable distributions and correlations across the near-optimal solutions. Furthermore, as identified in Ref. [30], all MGA scenarios studied include an extreme implementation of one or more technologies thereby diminishing the investment flexibility for the remaining technologies. Not considering solutions with technology compromises is far from ideal.

In this paper the novel numerical method MAA (Modeling All Alternatives) capable of determining the continuum of near-optimal solutions from a given energy system optimization model building on the principles of MGA is presented. The method allows for an exploration of all techno-economical near-optimal solutions, within a given slack on model objective value (system cost). By sampling the continuum of near-optimal solutions evenly, non-biased coverage of the alternative solutions is insured. Furthermore, analysis of technology correlations among alternatives are easily calculated. As the sampling process is computationally inexpensive, it is possible to obtain a dataset representing $500 \cdot 10^3$ near-optimal alternatives, thereby describing the entire continuum of near-optimal solutions well. Hereby moving the field of energy system optimization from discrete solutions to a new era where continuous solution ranges are available.

A model of the European electricity system [35], is used to validate the developed method. Using MAA the continuum of near-optimal solutions is identified, enabling us to consider dimensions usually neglected, such as time (speed of implementation), space (land use), national energy self-sufficiency, and economic equality of the solution.

This paper begins with an explanation of the MAA method, describing the mathematical formulation. It continues by introducing an application example based on a model of the European energy system. The results of applying the method to the example are thereafter presented and discussed. Finally a discussion of the presented method, its applications and limitations is presented.

## 2. Modeling All Alternatives

The type of optimization models studied in this work are linear models on the form given in Equation (2). Model parameters are contained in the matrices **A** and **C**, along with the cost vector **c**, and the vectors **d** and **b**.

$$\begin{aligned}\text{minimize} \quad & f(\mathbf{x}) = \mathbf{c} \cdot \mathbf{x} \\ \text{subject to} \quad & \mathbf{Cx} \leq \mathbf{d} \\ & \mathbf{Ax} = \mathbf{b};\end{aligned} \quad (1)$$

All solutions **x** satisfying the specified constraints are contained in the feasible space $X$:

$$X = \{\mathbf{x} \mid \mathbf{Cx} \leq \mathbf{d}, \mathbf{Ax} = \mathbf{b}\} \quad (2)$$

The goal of the MAA algorithm is to determine all solutions located near the model optimum **x*** within the solution space. The method developed can be divided into two phases, where the first phase is concerned with determining the polyhedron defining the near-optimal feasible space, and the second phase samples the polyhedron to create a dataset of samples that can be further studied. A schematic of the procedure is shown in Fig. 2.

### 2.1. Phase 1

Initially, the optimization problem is solved using the original objective function $f$ to find the economic optimum. This provides a single point **x*** located on the border of or within the feasible space. All solutions near the optimum **x*** are contained in a subspace $W$ of the feasible space $X$. Using the MGA constraint, formulated in Ref. [23], limiting the maximum increase in the original objective function value, the near-optimal solution space $W$ can be defined as:

$$W = \{\mathbf{x} | \mathbf{x} \in X, f(\mathbf{x}) \leq f(\mathbf{x}^*)(1+\varepsilon)\} \quad (3)$$

Here $\varepsilon$ is a user specified slack on the objective value. As all constraints defining $X$, and the MGA constraint $f(\mathbf{x}) \leq f(\mathbf{x}^\wedge)(1+\varepsilon)$ are linear, the near-optimal feasible space $W$ is a convex polyhedron as presented on Fig. 2. Because the near-optimal solution space $W$ is closed, any choice of objective function will provide a solution located on the perimeter of the space. Using a unit vector **n**, multiplied with the variables to be optimized **x**, as the objective function, allows for full control over the search/optimization direction given by **n**. By discarding the original objective function, introducing the MGA constraint and using the MAA objective function $f_{MAA}$, all vertices defining $W$ can be found be itteratively changing **n** and solving the optimization problem specified as:

$$\begin{aligned}\text{minimize} \quad & f_{\text{MAA}}(\mathbf{x}) = \mathbf{n} \cdot \mathbf{x} \\ \text{subject to} \quad & \mathbf{x} \in W\end{aligned} \quad (4)$$

Changing the model objective function to be on the form given in Equation (2) and selecting search directions **n** that seek to maximize and minimize every single variable in **x** one by one, as the MGA method presented in Ref. [30], a set of alternative solutions are found.

Knowing that the feasible region can be defined by a polyhedron, it makes sense to imitate this shape by computing the polyhedron containing all points found so far. Computation of the polyhedron containing the found alternative solutions are done with the quickhull algorithm [36]. Using the face normal vectors of the found polyhedron to define the next set of search directions **n**, as seen in Fig. 2, ensures that if one of the faces in the polyhedron is not part of the polyhedron defining the near-optimal feasible region, then a new point will be found when searching in the normal direction of that particular face. The newly found points combined with all previously found points, are used to repeat the process of defining a polyhedron and searching in the face normal directions. As long as the polyhedron computed doesn't describe the full near-optimal feasible region, new points will be found on the perimeter of the near-optimal feasible region, until all points defining the near-optimal feasible region are found. In other words, this method ensures that the polyhedron computed in every iteration converges towards the polyhedron defining the near-optimal solution space in a finite number of iterations.

If the near-optimal solution space $W$ was to have a very complex shape being defined by a high number of vertexes, making it computationally infeasible to fully compute, it would be necessary to have a termination criterion that does not require that the complete near-optimal feasible region is found. The volume of the polyhedron estimating the feasible region will converge towards the volume of the feasible region and implementing a convergence criteria on the volume provides a good termination criterion. Development in volume is represented in Fig. 2. The quickhull algorithm [36] is used to calculate polyhedron volume.

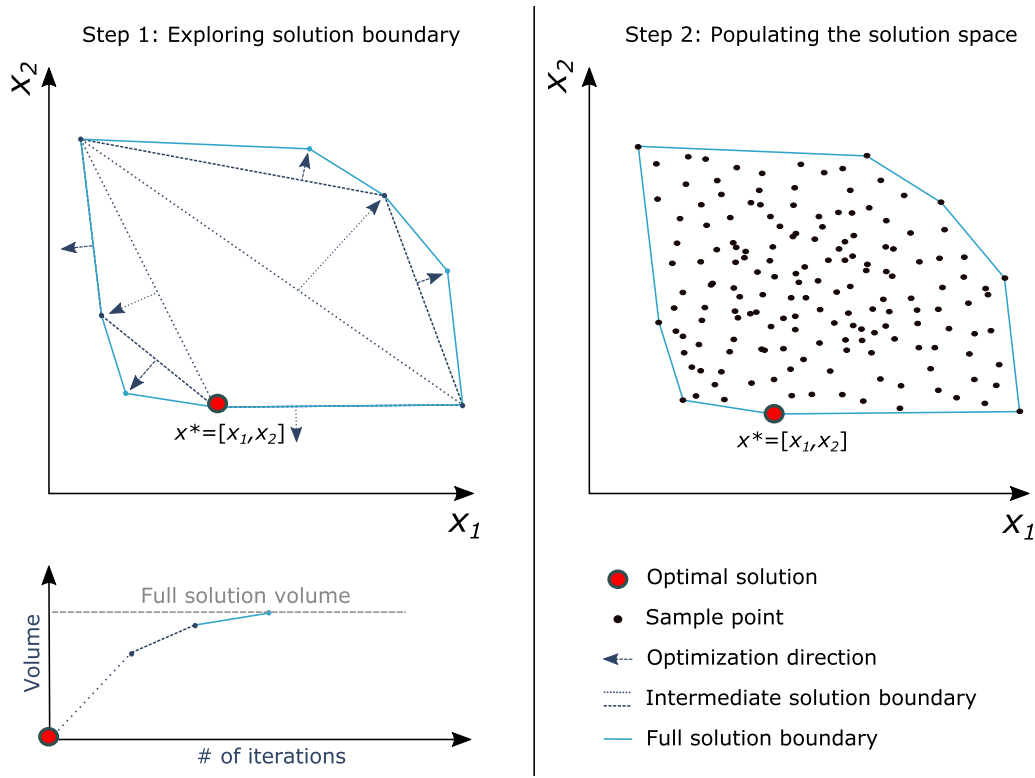

**Fig. 2.** A schematic of the MAA method - The left column displays step 1 in the MAA method. From the optimal solution the boundary of possibly optimal solutions is expanded by altering the objective function to the optimization problem. The volume of possible solutions is used as termination criteria, as it converges towards the volume of all possible solutions. In the right column of the figure, step 2 of the MAA method is displayed. Step 2 samples the bounded region to generate a dataset representing all possible optimal solutions to the problem.

### 2.2. Phase 2

Due to convexity, all solutions located within the polyhedron spanned by the extreme solutions are valid near-optimal solutions to the optimization problem [37]. In the second phase of the MAA method, the found near-optimal feasible space is sampled to create a well-representing dataset. More generally speaking the task is to draw samples evenly inside a polyhedron, which can be further reduced to simply drawing samples evenly spaced inside a simplex, as the polyhedron can be split into a range of simplexes. By simplex, it is the simplest geometrical shape spanning a volume in the given space that is refereed to. In 2-D space, this would be a triangle, 3-D space a tetrahedron, etc. Using the Qhull software [36] build on the quickhull algorithm, the convex hull containing all near-optimal feasible solutions, are split into simplexes. By drawing several samples equivalent to the volume fraction of the simplex multiplied with the number of total sample points desired, from each simplex, it is possible to sample the entire solution space evenly.

Each simplex is given by a list containing all its vertices $P = \{\mathbf{p}^1, \mathbf{p}^2, ..., \mathbf{p}^m\}$. The number of vertices needed to describe a simplex will always be $m = d + 1$, where $d$ describes the dimension of the space the simplex is located within. Any point inside a simplex can be described as a sum of the points describing the simplex $P$ scaled with a vector $\mathbf{s}$ if this scaling vector has the property of summing to one $\sum_i s_i = 1$:

$$\mathbf{p}_{\text{new}} = \sum_{i=1}^{m} \mathbf{p}^i s_i \tag{5}$$

Where $m$ is the number of points used to describe the simplex. Selecting $\mathbf{s}$, such that the space inside the simplex is sampled evenly is done using a method called the Bayesian Bootstrap [38].

Using the Bayesian Bootstrap method, an initial vector $\mathbf{r}$ containing $m - 1$ random component drawn, from an even distribution with a range from 0 to 1 is created. Then sorting the components of the vector $\mathbf{r}$ by increasing value, and adding 0 as the first entry and 1 as the last, this new $\mathbf{r}$ vector can be used to define a scaling vector. The length of this new $\mathbf{r}$ vector is now $m + 1$ as 0 and 1 has been added. Using the difference between the components in $\mathbf{r}$ to define a new vector:

$$\mathbf{s} = \{r_{i+1} - r_i\} \ \forall \ i = 1, 2, ..., m \tag{6}$$

The vector $\mathbf{s}$ has the property that the sum of the components will always be equal to 1, by definition. Using the $\mathbf{s}$ vector to scale the points in $P$ it is possible to draw a point randomly located within the simplex. Following this procedure for all simplexes provides an even sampling of the entire polyhedron.

Other approaches for sampling a convex polyhedron is available, such as the Hit-And-Run method presented in Ref. [39]. Alternatively one could also use MCMC (Markov Chain Monte Carlo) samplers such as a acceptance-rejection sampler [40]. These samplers could be used interchangeably without altering the method outcome significantly. The sampling method described in this section was chosen as it is capable of drawing more than $10^5$ samples pr. minute and is simple to implement.

*2.3. Limitations*

The hardware used in this experiment is a 64 core compute node with 500 GB of ram. With this setup, it was not feasible to use the Quickhull algorithm [36] in dimensions larger than 10. Because of the large number of variables in energy system optimization models, a set of derived variables are studied rather than the individual variables in the model. Inspired by the work [30] the derived variables consists of technology capacity sums for individual technologies, e.g. global wind power capacity, global solar PV capacity, etc. The groups of variables could, however, also be formed in alternative ways, e.g. considering only the technology capacities in a single country. In this paper, the grouping of variables provides several derived variables equal to the number of technologies included in the model.

The MAA method requires that the optimization problem is convex, thus making it unsuitable for mixed integer problems.

When studying the continuum of model solutions represented with derived variables, the density of the continuum varies across its extent, as some solutions can be achieved using a higher number of system configurations than others. This happens as a side effect, known as multiplicity, of simplifying the problem by summing model variables. Determining the variation in density is however very computationally intensive and is outside of the scope of this paper. Several approaches for determining the multiplicity are available, including Markov Chain Monte Carlo sampling or nested MAA iterations.

## 3. Application example

To verify the usefulness of the developed method, it has been tested on a model of the European electricity system presented in Ref. [35]. The optimization problem is formulated as a greenfield optimization problem and is used to determine the optimal power system configurations under several global $CO_2$ emissions constraints. Long-term market equilibrium, as well as perfect competition and foresight, are assumed. The model is built on the open framework PyPSA [6]. A brief description of the model will be given here, based on the thorough walk-through given in Ref. [35].

Using the MAA method a range of alternative solutions to the optimization problem will be identified. These solutions will be near the optimal solution in terms of total system cost, as a maximum increase of 10% is allowed. Throughout the results four distinct scenarios will be highlighted. These are the cost-optimal solution referred to as "optimum", the solution with highest equality in the distribution of weighted energy production referred to as "high equality", the solution performing best in terms of CO2 reduction referred to as "low $CO_2$ emission", and the solution with highest installation of wind power capacity referred to as "large wind capacity".

To compare the benefits and weaknesses of the MAA method compared to state-of-the-art MGA algorithms, a comparison study was performed. Using the MGA method presented in Ref. [30], the experiment was repeated for a global $CO_2$ reduction constraint of 95%.

*3.1. Model*

The model consists of a network of 30 nodes each representing a country of Europe with 50 cross border AC and DC transmission lines operated by members of the European Network of Transmission System Operators, ENTSO-E [41]. Hourly temporal data resolution for energy demand and wind/solar profiles for the specific nodes is used. A full year is simulated using the reference year 2011 for energy demand and wind/solar profiles as it provides a worst case scenario. Fig. 3 shows the network topology, which is based on currently installed international transmission lines. The model is formulated as a single objective linear optimization problem, modeling power flows using Kirchoff's law [42].

Each node in the network has energy-producing technologies available, with initial capacities being zero. The generators used in this project are onshore wind, offshore wind, solar PV, and open-cycle gas turbines (backup capacity). In the model, all technology capacities are expandable limited only by the geographical potential. Two storage technologies are included in the model. These are hydrogen storage and battery storage. Hydrogen storage serves as long term storage with a storage capacity large enough to store energy from 168 maximum discharge hours. The hydrogen storage is modeled as an electrolyzer/fuel-cell stack linked to a hydrogen storage tank, with a charging efficiency of 75% and discharge efficiency of 58%. The battery storage serves as a short term storage only capable of storing energy from 6 maximum discharge hours. The battery is modeled with a charging and discharging efficiency of 90%.

The model input data consists of technology costs provided for the individual technologies, specific hourly energy demand for the 30 model nodes, and specific hourly wind and solar availability in the model nodes. Technology costs are prediction for 2030 and are found in Table 1. The hourly energy production of all variable renewable energy sources is limited by the production potential given by the weather. Following [35], the availability was calculated using historical weather data [43]. The data for the hourly electricity demand found in the European Network of Transmission System Operators (ENTSOE) data portal is used as energy demand [44]. A total of 3152 TWh of energy was consumed by the countries combined in 2011. For a detailed description of the mathematical formulation of constraints and objective function refer to Ref. [35].

The model is solved to optimality using the commercial solver Gurobi [45] and the Barrier algorithm. When put on MPS format the model has ~ $7 \cdot 10^6$ rows, ~ $3 \cdot 10^6$ columns and ~ $14 \cdot 10^6$ nonzeros. It requires roughly 5 GB of memory and can be solved in less than 15 min on a 4 Core machine. When applying the MAA method the inequality constraint from equation (3) is added to the model, and the objective function is reformulated. This does not significantly alter the solving time.

*3.2. Socio-economic metrics*

To illustrate some of the possible applications of our method, a set of socio-economic variables is also considered. In this paper, the Gini coefficient has been used to measure national energy self-sufficiency and investment equality. A low Gini coefficient represents high self-sufficiency/equality and a high Gini coefficient means high dependence/inequality. Two further socio-economic metrics of the near-optimal solutions, i.e. land use and implementation time are also used to study the near-optimal solutions.

*3.2.1. Gini coefficient*

When modeling energy systems on the continental scale, it is important to consider how resources and workloads are distributed, as political incentives often favor an equal distribution, such that no single country or region has to carry a much larger burden than others. This is however not considered in most works related to techno-economic energy system optimization with models spanning multiple countries. The Gini coefficient has been used to calculate two equality measures. The first being equality in energy generation relative to demand [47]. This measure will be referred to as the national self-sufficiency. The second is equality in investment relative to energy demand.

To calculate the Gini coefficient representing national self-sufficiency, the cumulative share of demand per country is calculated and plotted against the cumulative share of generation per country. Thereby one gets the Lorenz curve for that specific solution, as shown with the blue line in Fig. 4. As inequality increases the Lorenz curve lies further and further away from the equality line, shown as the red line in Fig. 4.

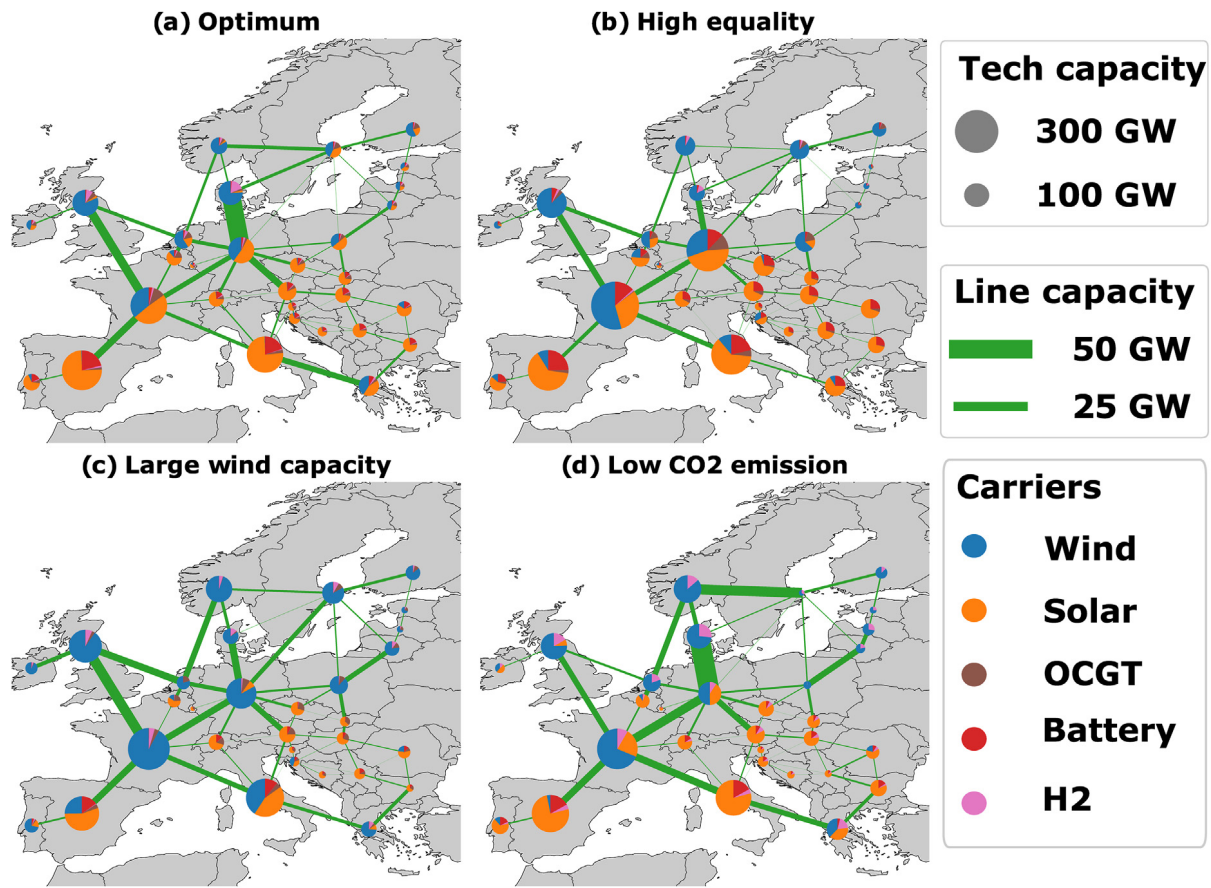

**Fig. 3.** Alternative near-optimal solutions to the PyPSA-Eur-30 model - Schematic of network topology from the PyPSA-Eur-30 model [35], used in this paper. The figure displays the optimal solution (a) as well as three alternative solutions (b) to (d) when a 95% $CO_2$ reduction constraint is enforced relative to 1990 emission values. Optimized installed capacities of energy generating technologies; wind turbines, solar PV, and open-cycle gas turbines (Backup) as well as storage units; Hydrogen storage ($H_2$), and battery storage are displayed in the individual nodes of the network. A ratio of energy to power capacity of 6 and 168 h is assumed for battery and hydrogen storage respectively. Optimized transmission line capacities are indicated by the line width.

**Table 1**
Technology cost data for all technologies included in the model. All values are predictions for 2030 from Ref. [46]. A discount rate of 7% is used to annualize costs. Storage has a two part capital cost as there is a cost for charging/discharging capacity listed in \euro; /MW, and a cost for storage capacity listed in \euro; /kWh. Transmission lines have an additional capital cost of 150 k \euro; pr. line, to cover the installation cost of transformation stations. OCGT is the only $CO_2$ emitting producing 0.19 t $CO_2$/MWh.

| Technology | Investement \euro; /kW | Fixed O&M \euro; /kW/year | Marginal cost \euro; /MWh | Lifetime years |
|---|---|---|---|---|
| Onshore Wind | 1035 | 12 | 0 | 30 |
| Offshore Wind | 1934 | 36 | 0 | 30 |
| Solar PV | 254 | 7 | 0 | 30 |
| OCGT | 435 | 7 | 58.4 | 25 |
| $H_2$ storage | 555 + 8.4 \euro; /kWh | 9.2 | 0 | 20 |
| Battery | 310 + 144 \euro; /kWh | 9.3 | 0 | 20 |
| Transmission | 400 \euro; /MW km +150 k \euro; pr. line | 2% | 0 | 40 |

The Gini coefficient is calculated as the ratio between the area enclosed by the Lorenz curve and the equality line (Area A on Fig. 4) relative to the total area under the equality line (Area A + B on Fig. 4). Thus, the Gini coefficient becomes $G = \frac{A}{A+B}$.

A scenario where every country over the duration of an entire year, produces as much energy as it consumes, would have a Gini coefficient of 0, and represent the equality line on Fig. 4. A scenario where one country is producing all energy, would, on the other hand, have a Gini coefficient of 1, and represent total inequality.

The Gini coefficient can also be modified to measure equality among other metrics, such as investment. By using the cumulative share of investment on the y-axis instead of the cumulative share of energy generation, a Gini coefficient representing equality in investment versus consumption is calculated.

*3.2.2. Land use*

Land use is calculated with energy density 20 MW/km$^2$ for onshore wind turbines, as the average turbine is set to have 5 MW capacity taking up a 500 × 500 m space. Offshore wind turbines are set to have zero land use. The energy density of solar PV plants used is 145 MW/km$^2$ as under reference conditions, the solar irradiance is 1000 W/m$^2$, 18% efficiency is assumed for the solar panels, and 20% extra space around the panels is considered for the installation. Gas turbines, H$_2$ storage, and battery storage are considered to have zero land use, as the plants are small relative to the variable renewable energy sources.

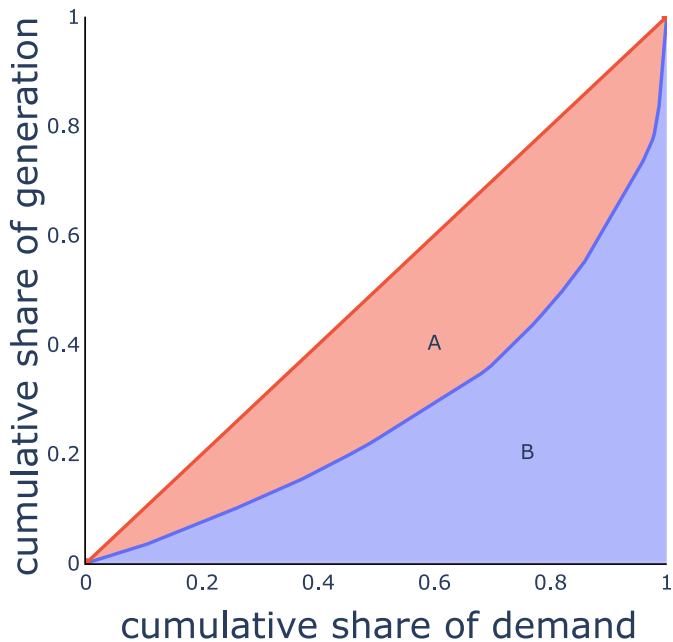

**Fig. 4. Energy demand Lorentz curve** - The Figure shows the Lorentz curve (blue line) for the energy demand/generation in a given scenario. The equality line is shown as the red line. The two areas A and B used to calculate the Gini coefficient is shown on the figure. (For interpretation of the references to colour in this figure legend, the reader is referred to the Web version of this article.)

*3.2.3. Implementation time*

Implementation time of the near-optimal solutions is calculated as the time it would take to implement a scenario, in the slowest country, if all countries are restricted to spend no more than 10% of their GDP on energy system renewal per year, transmission not included. January 2020 GDP values, calculated by the World Bank, were used.

**4. Results**

The MAA method has been applied to the model described while enforcing a 95% CO$_2$ reduction constraint compared to 1990 emissions and allowing for a 10% increase in system cost relative to the cheapest solutions with similar CO$_2$ reduction constraint. Phase 1 of the MAA algorithm converged after obtaining 244 near-optimal solutions. Sampling the near-optimal solution space 500 · 10$^3$ times, in Phase 2 of the MAA algorithm, yields the results presented in Fig. 5. Variable distributions are displayed on the diagonal, with pairwise Pearson correlation values printed on the top-right side of the figure, and contour plots of variable correlations on the lower-left side. Four sample solutions each representing an extreme in the dimensions: minimum system cost, equality in production, wind capacity implementation, and CO$_2$ reduction beyond the 95% requirement, are shown in Fig. 3.

Studying variable distributions on the diagonal of Fig. 5, solar PV, and wind capacities are seen covering ranges wider than 1 TW. The wide spread in variable values indicates that the model optimum is relatively flat, and therefore allows for large variations in model solutions with small changes in model objective (system cost). The flat nature of the model optimum underlines the importance of analyzing near-optimal solutions.

Throughout the plots in Fig. 5, the four scenarios (Optimum, High Equality, Large wind turbine capacity, Low CO$_2$ emission) from Fig. 3 are shown. By analyzing a single scenario in Fig. 5, it is possible to see how a choice in one variable affects the allowable ranges in other variables. An example would be if high national self-sufficiency in energy production is desired (low Gini coefficient), then by analyzing the Gini coefficient versus backup capacity plot in Fig. 5, it can be seen that requiring a Gini coefficient below 0.15 constraints the amount of backup capacity to be around 150 GW. Furthermore, by analyzing the Hydrogen storage versus Gini coefficient plot, it is seen that a Hydrogen storage capacity is decreased slightly from the optimal solution to approximately 50 GW, to achieve the low Gini coefficient of 0.15. Essentially Fig. 5 serves as a tool allowing for decision-makers to design an optimal solution satisfying as many unmodeled objectives and constraints as possible, without having to select between discrete scenarios.

Studying variable correlations on the upper right half of Fig. 5, a strong negative correlation between wind and solar power of −0.6 is seen. As these two technologies are the only renewable energy sources included in the model, they are directly competing, thus such strong negative correlations are to be expected. Analyzing the correlations of wind and solar power with the Gini-coefficient representing national self-sufficiency, a strong negative correlation is seen for solar power, and a significantly lower correlation is seen for wind power. Remembering that a low Gini coefficient represents a high level of national self-sufficiency, the strong negative correlation seen for solar power indicates that increasing solar power capacity increases national self-sufficiency. This corresponds well with results from literature where local installations of solar power plants are seen as the solar availability time series have been found to correlate on a large spatial scale across Europe. Studying the role of hydrogen storage, a strong negative correlation is seen with OCGT backup capacity, indicating that hydrogen storage is directly competing with OCGT to provide energy in periods of scarce renewable energy sources. The short-term battery storage, on the other hand, has no significant negative correlation with OCGT backup capacity but does however have strong correlations with solar power.

Applying the MAA method to a series of scenarios with an increasing constraint on CO$_2$ emission provides information about the possible variations in model solutions in the transition of an energy system. The data used in Fig. 6 was generated by using the MAA method at four CO$_2$ reduction scenarios (unconstrained, 50, 80, and 95%). Three levels of slack on total system cost were used, 15, 30, and 45% calculated relative to the optimal solution without any constraint on CO$_2$ emissions. As the only non-renewable energy source included in the model is open cycle gas turbines which are most suitable as backup generation, the unconstrained and 50% reduction scenario has a reduction in CO$_2$ emissions above 60% relative to 1990 values. As a result, the lowest-cost scenario for this model has a CO$_2$ emission reduction of 62% as seen in Fig. 6(**f**) Since the interest of this paper is highly renewable energy futures with large CO$_2$ reductions, this behavior is accepted. Analyzing the Gini coefficient representing national self-sufficiency, it is seen how the Gini coefficient increases, indicating lower self-sufficiency, as CO$_2$ emissions are reduced. This effect is seen as the cost optimization install renewable energy sources in locations with favorable resources, rather than where the power is needed when CO$_2$ emissions are reduced. Fig. 6(**a**) does, however, reveal a lot of flexibility allowing for higher self-sufficiency at small increases in total system cost until CO$_2$ emission reductions surpass 95%. Rising trends are, furthermore, seen for the remaining metrics on Fig. 6(**b-e**), combined with rising system cost, 6(**f**). Using Fig. 6, decision-makers can design alternative transition pathways, late and rapid, or early and steady, as studied by Marta Victoria [11], towards a decarbonized energy system, without having to perform time-consuming modeling work. The figure allows decision-makers to take measures in the early stages to prevent undesirable developments such as increasing inequality.

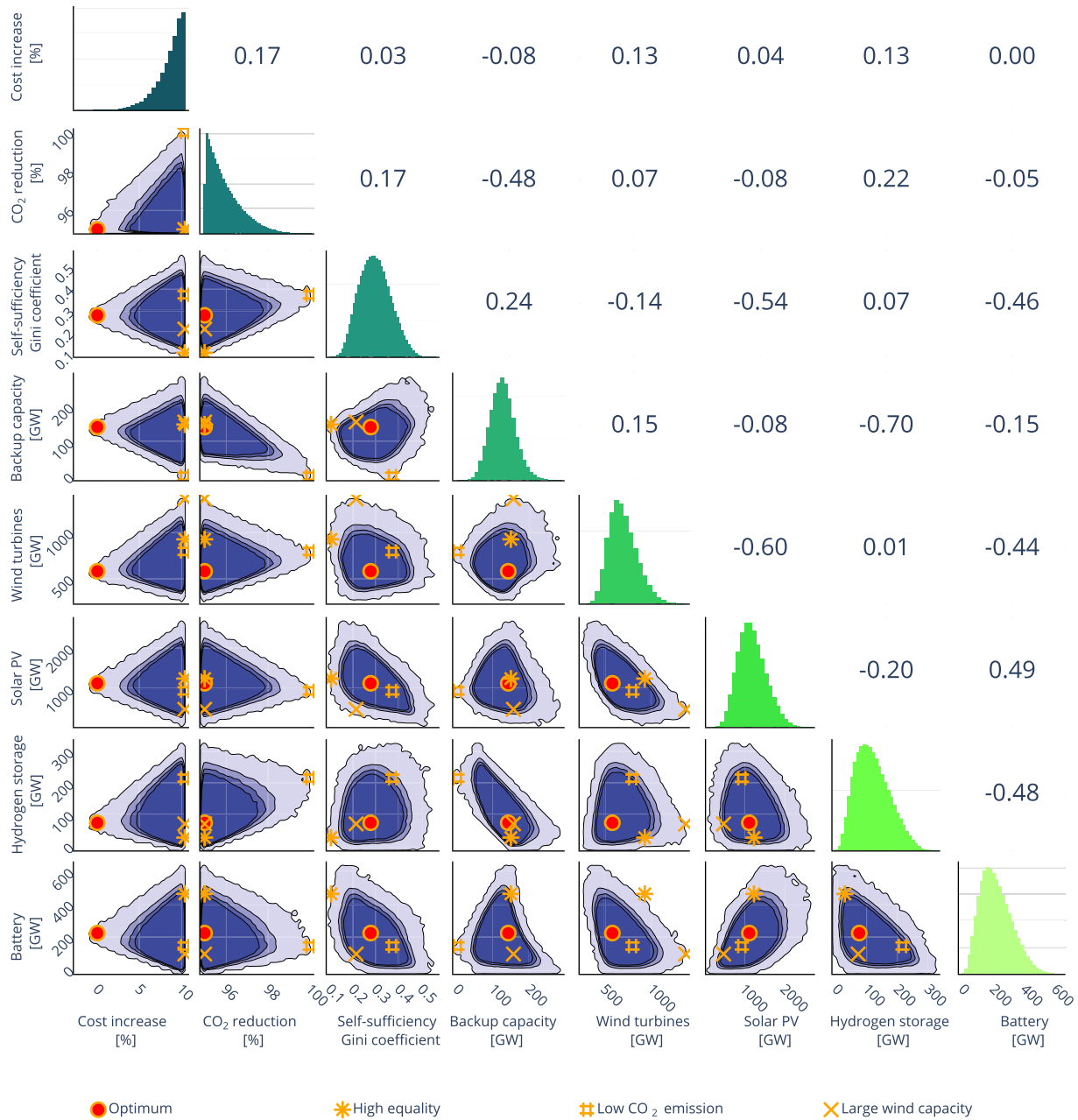

**Fig. 5.** Variable correlations amongst all near-optimal solutions in a 95% CO₂ reduction scenario - On the figure diagonal normalized variable distributions are shown. Variable correlations are shown on the top-right side of the diagonal. In the lower-left side, contour plots revealing density in solutions are shown. The variations in density arises from collapsing high dimensional data to 2-D figures. The techno-economical optimal solution is marked with the red dot. Furthermore, the three scenarios shown in Fig. 3; High equality, Low CO₂ emissions and large wind capacity are marked on the contour plots. A 95% CO₂ reduction constraint is enforced relative to 1990 emissions, combined with an allowable increase in system cost of 10% relative to the optimal solution. (For interpretation of the references to colour in this figure legend, the reader is referred to the Web version of this article.)

## 5. Discussion

The results found using the presented MAA method reveal large variations among model solutions, at small variations in model objective (total system cost). These findings indicate that the model optimum is relatively flat, causing structural flaws in the model to have a large effect on model output. Looking at the results from a decision-maker's perspective, the results reveal that large flexibility is available at only small changes in total system cost. In Ref. [48] a comprehensive review shows that only a limited number of studies based on ESOMs use systematic approaches to uncertainty quantification. Xiufeng Yue [48], further found that the majority of studies implement scenario analysis studying a small ensemble of alternatives centered around a base solution. The MAA method provides a robust framework providing identification of uncertainty ranges and variable correlations.

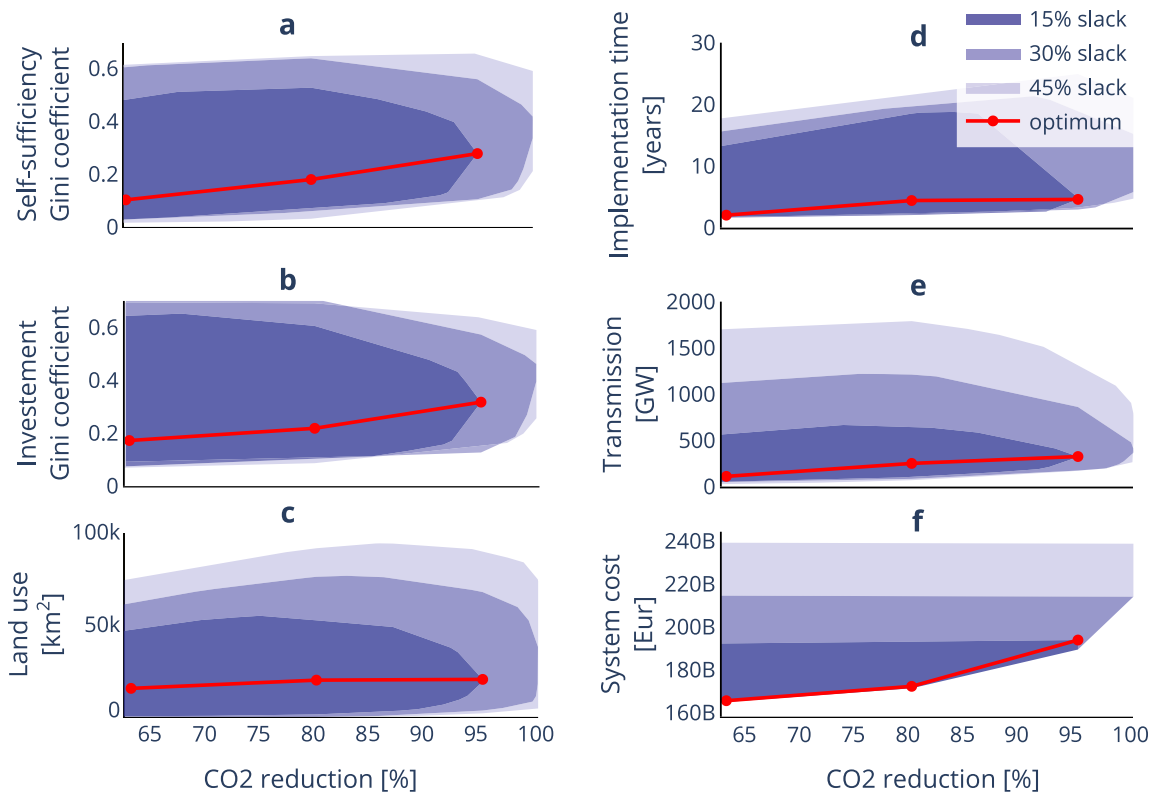

**Fig. 6. Variable distributions across a CO$_2$ reduction range** - Figure **a** to **e** shows a range of socio- and techno-economic metrics plotted against CO$_2$ emission reductions. Nine runs with the CO$_2$ reduction constraints 0,80 and 95% and slack values 15, 30 and 45%. The slack on total system cost is calculated relative to the overall cheapest solution of the nine runs. In Figure **f** the total system cost for the three levels of slack on objective value are shown. As the only emitting technology included in the model is OCGT, the scenario with unconstrained CO$_2$ emissions achieves a reduction of 63% compared to 1990 values.

Current state-of-the-art energy system studies often rely on scenario-based studies to identify possible futures. Using scenario-based optimization introduces bias as the modeler is responsible for selecting a diverse set of scenarios uncovering future possibilities. This bias is, however, removed when all possible solutions are identified with the MAA method, ensuring even coverage of future possibilities.

As recognized by Joseph DeCarlois [25], the HSJ and other MGA methods have no way of providing information about the robustness of found alternative solutions. As the MAA method identifies the entire set of near-optimal solutions within a given slack, information about variable robustness is implicitly identified. Analyzing variable distributions on the diagonal of Fig. 5, it is seen that the density varies across the variable range. The density is a measure of how many valid configurations of the model are achievable with a given variable value. Selecting a scenario with variable values having a high density provides a robust solution, hence it can be achieved in several ways.

When comparing the results found using the MAA method to results found using the one at the time MGA approach from Ref. [30], presented in Fig. 7, it is clear that vast regions of the near-optimal solution space are left unexplored when MGA methods are used. By analyzing the cost increase of the MGA and MAA solutions on Fig. 7 in the first columns, it is clear that all MGA solutions utilize the entire cost slack allowed. Thereby not providing information about intermediate solutions. Furthermore, looking at the wind capacity vs. hydrogen storage axes in Fig. 7, its is clear that the MGA method identifies most extreme solutions and a few solutions around the optimum, but fail in providing a complete picture of the near-optimal boundary. These results, support the conclusion made in Ref. [33], that global approaches towards identifying uncertainties are needed rather than one at the time approaches if the entire solution space is to be discovered. Looking at the MAA results, the density of solutions is much higher in the region around the optimum and quite low at the extremes. The increased density of solutions indicate that solutions located here are more likely to be implemented as they can be achieved under several configurations of the energy network. As the majority of solutions found using MGA are located on the very boundary of the near-optimal solutions space where solution density is low, these solutions can be hard to implement as they leave little room for deviations. Using the MAA problem on this specific problem required solving the optimization problem 244 times, whereas the MGA method only required solving the problem 12 times. The additional computational cost can make the MAA inapplicable to very detailed energy system models requiring several hours to solve.

As the only requirement for using the developed MAA method on an optimization problem is convexity, the method can easily be adapted to other fields than energy system modeling. Applications such as logistics, water management, and public planning, where traditional robust and stochastic optimization approaches are used, would be applicable for the MAA method.

In literature, large variations in results from studies analyzing similar energy systems are seen. As the region around the model optimum is relatively flat for the model used in this work, one can hypothesize that this is a general feature found in energy system models. A flat optimal region can lead to large variations in results with small changes in objective function formulation. Objective functions subject to large uncertainties arising from real-world complexity, combined with the flat optimum found in energy system models, is a likely explanation for the large variations in results found in literature. All of those results may be near-optimal, but it is difficult to assess how robust they are and what other solutions exist out of the narrow view of classic optimization approaches. The MAA method presented in this paper avoids that limitation by providing all possible solutions and an indication of their probability, i.e., the MAA method widens our field of view when investigating energy systems and enables the simultaneous evaluation of different metrics.

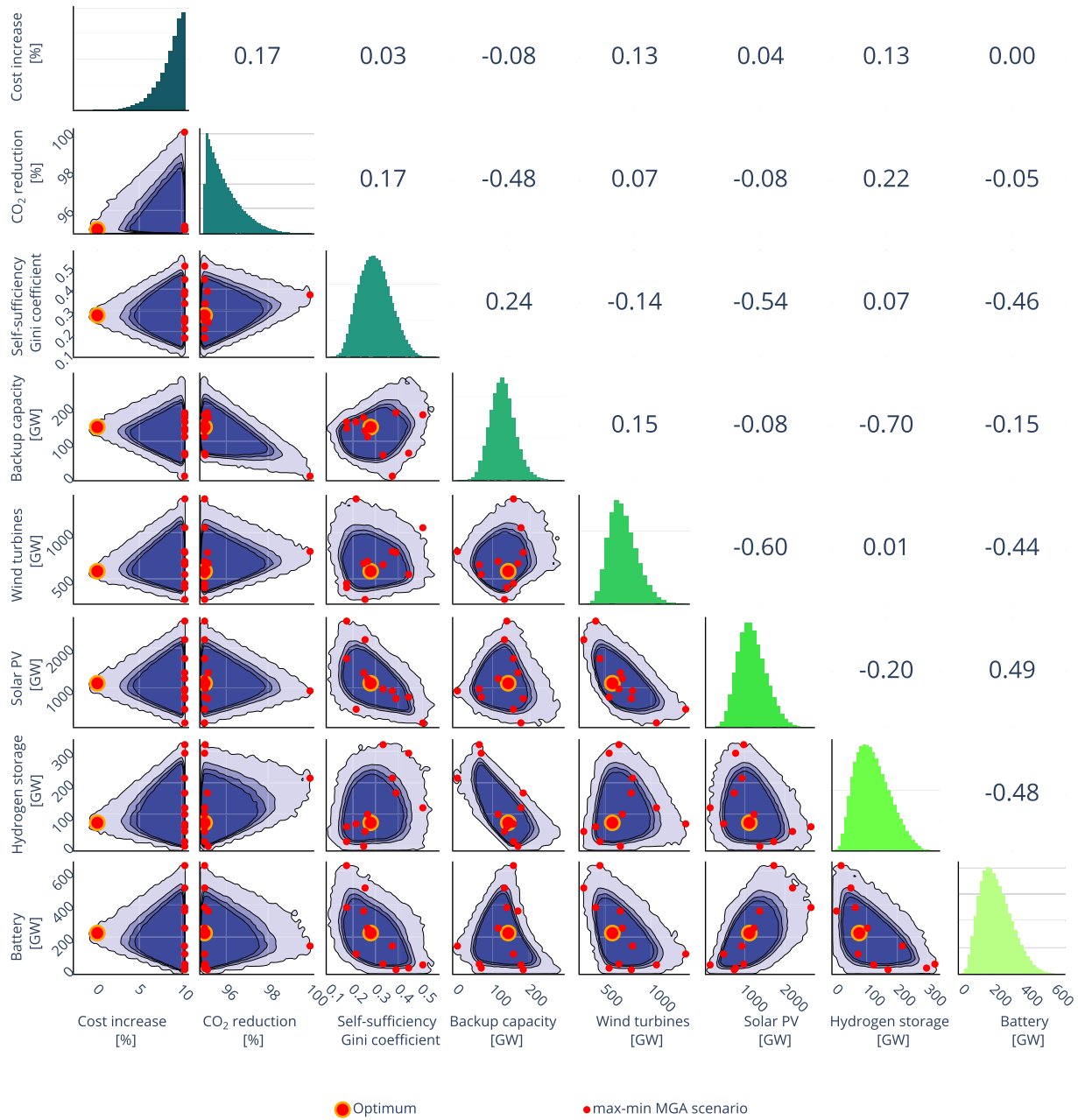

**Fig. 7. Comparison between MGA and MAA in a 95% CO₂ reduction scenario** - On the figure diagonal normalized variable distributions found using the MAA method are shown. Variable correlations are shown on the top-right side of the diagonal. In the lower-left side, contour plots revealing density in solutions are shown. Scenarios found using the max-min MGA method from Ref. [30] are shown as red dots. A 95% $CO_2$ reduction constraint is enforced relative to 1990 emissions, combined with an allowable increase in system cost of 10% compared to the cheapest solution.

## 6. Conclusion

In this paper, a method capable of identifying all near-optimal solutions to energy system optimization models is proposed. The method capabilities are explored by the application of the method on a model of the European energy supply.

Initially, the current state-of-the-art MGA methods applied in energy system optimization is considered. Two core problems are identified namely; a) no guarantee for uniform coverage of the near-optimal feasible space and b) too few, and extreme alternatives are identified, thereby not allowing for analysis of technology correlations and study of intermediate solutions.

Building on the principles of previous MGA methods the novel

MAA method is proposed. The proposed method distinguishes itself from current state-of-the-art methods on two accounts. a) it ensures convergence towards complete coverage of the near-optimal feasible space of a given optimization model. b) By sampling intermediate solutions from the near-optimal feasible space, the number of alternative solutions considered are increased with several orders of magnitude, going from the range of $10^2$ alternatives to $10^5$.

Applying the proposed MAA method on a model of the European energy supply reveals large variations among alternative solutions at small variations in total system cost. The large variations indicate that it is indeed naive to study a single optimal or a handful of near-optimal solutions, as the large solution flexibility will not be identified.

Using the large number of alternative solutions identified, technology correlations are determined and analyzed. Using technology correlations, expected results such as a strong correlation between solar PV capacity and battery storage are identified. Furthermore, hydrogen storage is found to have a strong negative correlation with OCGT backup capacity, indicating that these two technologies both can serve as a backup resource in periods of scarce availability of renewable energy sources. To further establish the usefulness of the proposed method, a series of increasing $CO_2$ reduction scenarios are studied. Again, large variations among alternative solutions are found, indicating large flexibility in the design of transition paths for the European energy supply.

## Software availability

Name: Modeling All Alternatives.

## Description

The developed python scripts used in the paper are available under an open-source license. All scripts and data required to reproduce the results from this paper are openly available.

## License

The software is under the GNUv3 license.

## Resources

Https://github.com/TimToernes/ModelingAllAlternatives.

## Requirements

- Python 3.7
- Compute node with 32 cores and 256G ram

## Credit author statement

Tim T. Pedersen: Methodology, Software, Writing – Original Draft, Data Curation, Marta Victoria: Validation, Resources, Writing – Review & Editing, Supervision, Morten G Rasmussen: Conceptualization, Methodology, Gorm B. Andresen: Conceptualization, Writing – Review & Editing, Supervision, Funding acquisition.

## Declaration of competing interest

The authors declare that they have no known competing financial interests or personal relationships that could have appeared to influence the work reported in this paper.

## Acknowledgements

Tim T. Pedersen, M. Victoria nad Gorm B. Andresen are fully or partially funded by the RE-INVEST project, which is supported by the Innovation Fund Denmark under grant number 6154-00022B. Tim T. Pedersen is fully or partially funded by the FIRE project, which is supported by the European Regional Development Fund under grant number RFD-16-0024. The responsibility for the content lies solely with the authors.